\documentclass[10pt,conference]{IEEEtran}
\IEEEoverridecommandlockouts

\usepackage{tikz}
\usepackage{amsmath}
\usepackage{amsfonts}
\usepackage{textcomp}
\usepackage[table]{xcolor}
\usepackage{url}
\usepackage{afterpage}
\usepackage{subcaption}
\usepackage{hyperref}
\usepackage{listings}
\usepackage{booktabs}
\usepackage{algorithm}
\usepackage{graphicx}
\usepackage{fancybox}
\usepackage{xspace}
\usepackage{threeparttable}
\usepackage{multirow}
\usepackage{seqsplit}
\usepackage{caption}
\usepackage{subcaption}
\usepackage{pifont}
\usepackage{enumitem}
\usepackage{balance}
\usepackage{array} 
\usepackage{microtype}
\usepackage{booktabs}
\usepackage{tabularx}
\usepackage{multirow}
\usepackage{algpseudocode}

\def\BibTeX{{\rm B\kern-.05em{\sc i\kern-.025em b}\kern-.08em
    T\kern-.1667em\lower.7ex\hbox{E}\kern-.125emX}}

\lstdefinestyle{BaseListing}{
  basicstyle=\scriptsize\ttfamily,
  columns=fullflexible,
  keepspaces=true,
  breaklines=true,
  breakatwhitespace=false,
  breakindent=1.5em,
  breakautoindent=true,
  showstringspaces=false,
  tabsize=2,
  frame=single,
  framerule=0.3pt,
  aboveskip=4pt,
  belowskip=4pt,
  xleftmargin=0.5em,
  xrightmargin=0.5em,
  lineskip=0pt
}

\definecolor{agentIRKey}{HTML}{781F7A}
\definecolor{agentIRType}{HTML}{004D99}
\definecolor{agentIRRel}{HTML}{047857}
\definecolor{agentIRComment}{HTML}{6B7280}
\definecolor{agentIRString}{HTML}{B45309}
\definecolor{agentIRBg}{HTML}{F8FAFC}

\lstdefinelanguage{AgentIR}{
  sensitive=true,
  alsoletter={_},
  comment=[l]{\#},
  morecomment=[l]{//},
  commentstyle=\color{agentIRComment}\itshape,
  morekeywords={
    AgentIR,facts,relations,common,fields,
    ExecutionUnit,Controller,Invocation,
    StateUpdate,Bound,ExitRecord
  },
  keywordstyle=\color{agentIRKey}\bfseries,
  morekeywords=[3]{
    set,list,dict,str,bool,int,float,Any,None
  },
  keywordstyle=[3]\color{agentIRRel}\bfseries,
  morestring=[b]',
  morestring=[b]",
  stringstyle=\color{agentIRString}
}

\lstdefinestyle{AgentIRSchema}{
  style=BaseListing,
  language=AgentIR,
  backgroundcolor=\color{agentIRBg},
  rulecolor=\color{agentIRComment},
  postbreak=\mbox{\textcolor{agentIRComment}{$\hookrightarrow$}\space}
}

\definecolor{agentIRId}{HTML}{C81E1E}    
\definecolor{agentIRConst}{HTML}{047857} 
\definecolor{agentIRBool}{HTML}{7C2D12}  

\lstdefinestyle{AgentIRExample}{
  style=AgentIRSchema,
  morekeywords=[4]{
    U0,C0,C1,X0,X1,S0,S1,T0,T1,B0,B1
  },
  keywordstyle=[4]\color{agentIRId}\bfseries,
  morekeywords=[5]{
    BASIC_BLOCK,LOOP,ROUTER,RETRY,TERMINATION,
    LLM_CALL,TOOL_CALL,AGENT_RUN,WORKFLOW_RUN,SUBPROCESS,
    STATE_APPEND,MEMORY_WRITE,FILE_WRITE,WORKFLOW_STATE_UPDATE,
    CONDITIONAL_FALSE,BREAK,RETURN,RAISE,TERMINATION_PREDICATE,
    Covered,UncoveredOrWeak,BoundStatus
  },
  keywordstyle=[5]\color{agentIRConst}\bfseries,
  morekeywords=[6]{
    true,false
  },
  keywordstyle=[6]\color{agentIRBool}\bfseries
}

\definecolor{aldgTop}{HTML}{B45309}   
\definecolor{aldgGroup}{HTML}{7C2D12} 
\definecolor{aldgConst}{HTML}{047857} 

\lstdefinestyle{ALDGKinds}{
  style=AgentIRSchema,
  morekeywords=[4]{
    ALDGNodes,ALDGEdges
  },
  keywordstyle=[3]\color{aldgTop}\bfseries,
  morekeywords=[4]{
    Scope,Controller,Invocation,StateUpdate,
    Structural,Exit,Framework,Feedback
  }
}

\definecolor{best}{HTML}{dff5f2}  
\definecolor{second}{HTML}{E3F2FD}

\newcommand{\toolname}{\textsc{IAL-Scan}}

\begin{document}

\title{When Agents Do Not Stop: Uncovering Infinite Agentic Loops in LLM Agents}

\author{
\IEEEauthorblockN{Xinyi Hou, Shenao Wang, Yanjie Zhao, and Haoyu Wang\IEEEauthorrefmark{1}}
\IEEEauthorblockA{
Huazhong University of Science and Technology, Wuhan, China\\
xinyihou@hust.edu.cn, shenaowang@hust.edu.cn, yanjie\_zhao@hust.edu.cn, haoyuwang@hust.edu.cn}
\thanks{\IEEEauthorrefmark{1}Haoyu Wang is the corresponding author (haoyuwang@hust.edu.cn).}
}

\maketitle

\begin{abstract}
LLM agents increasingly rely on iterative execution to solve tasks through planning, tool use, state updates, and agent collaboration. While this design enables flexible automation, it also creates a new class of failures: an agent may repeatedly execute model calls, tools, workflow transitions, or agent handoffs when the feedback path is not effectively bounded. We call this problem \emph{Infinite Agentic Loops} (IALs). IALs are not ordinary programming loops; they arise from the interaction between agent logic, framework semantics, runtime observations, and termination mechanisms. Such failures can amplify a single request into long running model and tool execution, causing cost exhaustion, model denial of service, context growth, and repeated external side effects.
We propose \toolname{}, a static analysis tool for detecting IAL failures in real-world LLM agent projects. \toolname{} abstracts heterogeneous agent code into a framework independent Agent IR, builds an Agentic Loop Dependence Graph (ALDG) to recover explicit and framework induced feedback paths, and checks whether these paths can repeatedly reach costly or state growing operations without an effective bound. 
We evaluate \toolname{} on 6,549 LLM agent repositories. It reports 74 potential findings, among which manual review confirms 68 IAL failures across 47 projects, achieving 91.9\% precision. 
\end{abstract}


\section{Introduction}
\label{sec:introduction}

Large language model (LLM) applications are rapidly evolving from chat-style applications~\cite{hou2025llmapp,zhao2025llmapp} into autonomous agents that operate through iterative perceive, reason, and act loops. One of the most representative agent paradigms is ReAct, in which agents iteratively reason and act based on external observations~\cite{yao2022react}. This paradigm has been widely adopted in modern agent frameworks such as AutoGen~\cite{wu2023autogen}, LangGraph~\cite{langgraph_recursion_limit}, LangChain~\cite{langchain_agentexecutor_max_iterations}, CrewAI~\cite{crewai_customizing_agents}, and the OpenAI Agents SDK~\cite{openai_agents_runner}. Across these systems, iterative execution has become a core feature, enabling agents to repeatedly reason, act, observe, and decide what to do next.

However, this iterative execution model also exposes LLM agents to a structural failure mode known as \emph{\textbf{Infinite Agentic Loops}} (IALs). We define an IAL as an execution failure in which an agentic feedback path repeatedly triggers LLM calls, tool invocations, agent executions, or workflow transitions without an effective termination condition. Such paths can arise from explicit loops, recursive calls, workflow cycles, retry or repair logic, tool reentry, or multi-agent delegation. Modern agent frameworks have explicitly supported iterative termination mechanisms, including LangChain's \texttt{max\_iterations}~\cite{langchain_agentexecutor_max_iterations}, LangGraph's recursion limits~\cite{langgraph_recursion_limit,langgraph_graph_recursion_error}, the OpenAI Agents SDK's maximum turn limit~\cite{openai_agents_runner,openai_agents_running}, and CrewAI's \texttt{max\_iter}~\cite{crewai_customizing_agents}. However, these mechanisms do not eliminate IAL risks in practice. Developers may omit them, misuse them, configure them with ineffective bounds, or place them outside the actual feedback path. Moreover, some termination conditions are semantically fragile because continuation may still be controlled by model outputs, tool observations, external state, exceptions, or delegation decisions. Recent blogs and community issues further show that IAL failures occur in real deployments~\cite{reddit_langgraph_loop,crewai_issue_330,langgraph_issue_6731}. As a result, IALs remain a practical risk in real agent programs, even when frameworks provide loop-control mechanisms.


Despite this risk, existing approaches do not adequately detect IALs before deployment. General static analysis tools such as CodeQL and Semgrep can find many source-level vulnerability patterns~\cite{codeql,semgrep}, but they do not model agent execution semantics such as framework API edges, tool dispatch, handoffs, and agent reentry. Recent agent analyses study workflow topology, tool effects, privileges, information flow, and audit evidence~\cite{xavier2026agentproof,zhang2026agentaudit,adam2026toolsafety,li2026agentbom,zhang2024agent,debenedetti2024agentdojo}. However, IAL detection requires checking whether a repeated agentic feedback path can keep reaching costly or state-growing actions, and whether an effective bound covers that path. This gap motivates a dedicated analysis for IAL failures.

Detecting IALs statically is challenging for three reasons. (1) Agent behavior is often encoded through framework interfaces rather than direct source-level calls. The same execution concepts, such as model invocation and tool dispatch, may appear through different APIs and configuration fields across frameworks. A detector therefore needs a common representation of framework-induced agent semantics instead of relying on syntactic loops or framework-specific API names. (2) An IAL is usually formed by a feedback path that spans multiple program and framework elements. The repeated path may connect model calls, routing predicates, tool branches, message or state updates, retry handlers, and agent handoffs, so the analysis must reconstruct control, call, and state relations across wrapper code and framework logic. (3) Detecting IALs requires reasoning about bound coverage rather than merely checking whether a limit or exit condition exists. Loops are common and often legitimate in agent applications, but they become unsafe when a feedback path can repeatedly trigger costly or state-growing operations without an effective bound that constrains the controller and covers the repeated path.

To address these challenges, we present \toolname{}, a static analyzer for detecting IAL failures in downstream LLM agent projects. \toolname{} first abstracts source code and framework behavior into a framework-independent Agent IR, which captures the execution elements needed to reason about agent loops. It then constructs an Agentic Loop Dependence Graph (ALDG) to recover both explicit loops and framework-induced feedback paths, such as workflow transitions, tool dispatch, retries, and agent reentry. Based on the ALDG, \toolname{} checks whether a reachable feedback path can repeatedly trigger costly or state-growing actions, how its continuation is controlled, and whether the path is covered by an effective bound. This paper makes the following contributions:

\begin{itemize}
    \item \textbf{Definition of IALs.}
    We define \emph{Infinite Agentic Loops} (IALs) as a structural execution failure where an agentic feedback path repeatedly triggers model, tool, agent, or workflow execution without an effective stopping bound. We distinguish IALs from legitimate agent iteration and characterize their causes and security impacts.

    \item \textbf{IAL Detection.}
    We design and implement \toolname{}, a static analyzer for IAL failures in downstream LLM agent projects. \toolname{} supports eight mainstream agent frameworks and identifies feedback paths through Agent IR and ALDG-based analysis. 

    \item \textbf{Real-world Findings.}
    We evaluate \toolname{} on 6,549 real-world LLM agent repositories. It reports 74 potential findings, of which 68 are confirmed IAL failures across 47 agent projects, yielding a precision of 91.9\%.
\end{itemize}
\section{Background and Related Work}
\label{sec:background}

\subsection{LLM Agents and Execution Loops}
\label{sec:bg-execution-loops}

\begin{figure}[t]
    \centering
    \includegraphics[width=1\linewidth]{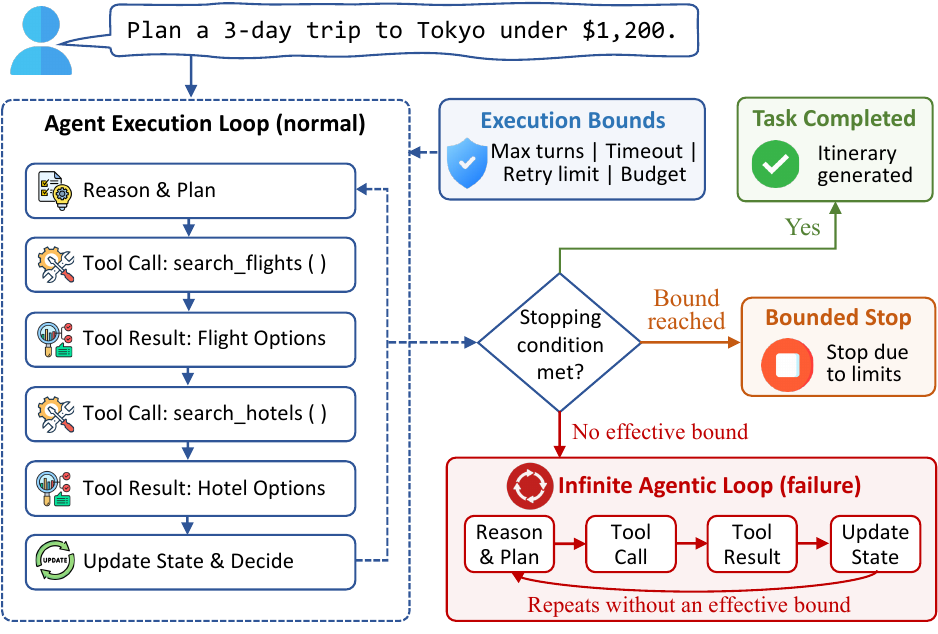}
    \caption{Execution loop of an LLM agent.}
    \label{fig:background}
\end{figure}

LLM agents extend conventional LLM applications from direct text generation to systems that plan, call tools, observe results, update state, and decide subsequent actions. This execution model is reflected in ReAct-style reasoning and acting~\cite{yao2023react}, tool-use methods such as Toolformer~\cite{schick2023toolformer}, feedback-based agents such as Reflexion~\cite{shinn2023reflexion}, and multi-agent systems such as AutoGen~\cite{wu2023autogen}. Recent surveys also identify planning, memory, tool use, and action as core components of LLM agent systems~\cite{wang2023survey,li2024personal,wang2025largemodelbasedagents}.
These capabilities are often implemented through agent frameworks such as LangChain~\cite{langchain_home}, LangGraph~\cite{langgraph_overview}, OpenAI Agents SDK~\cite{openai_agents_sdk}, AutoGen~\cite{autogen_update}, and CrewAI~\cite{crewai_github}. 
\autoref{fig:background} illustrates this execution loop. Given a user task, the agent repeatedly reasons, calls tools, observes results, and updates state until the task completes or an execution bound is reached. Such bounds include maximum turns, timeouts, retry limits, budgets, and recursion limits. If no effective bound covers the feedback path, the same loop can degrade into an infinite agentic loop that repeatedly triggers model calls, tool calls, state updates, or agent transitions.

Mainstream frameworks already expose such bounds, confirming that loop control is a practical concern. LangChain provides \texttt{max\_iterations} and warns that disabling it may cause infinite loops~\cite{langchain_agentexecutor}. The OpenAI Agents SDK uses \texttt{max\_turns} and raises \texttt{MaxTurnsExceeded} when the limit is exceeded~\cite{openai_running_agents,openai_runner_ref}. AutoGen, LangGraph, and CrewAI similarly provide termination conditions, recursion limits, or iteration caps~\cite{autogen_termination,langgraph_recursion_limit,crewai_customizing_agents}. These mechanisms show that the key issue is not the presence of a loop, but whether an effective bound covers its feedback path.

\subsection{Static Analysis for LLM Agents}
\label{sec:bg-static-analysis-agents}

Recent studies have begun to analyze LLM agent systems before or during deployment. AgentProof verifies structural properties of extracted workflow graphs~\cite{xavier2026agentproof}. Agent Audit combines dataflow analysis, credential detection, configuration parsing, and privilege checks for Python agent applications~\cite{zhang2026agentaudit}. Other work studies tool effects and information flow: Adam et al. statically summarize tool effects and check sandbox usage~\cite{adam2026toolsafety}; AgentRaft builds cross tool function call graphs for data over exposure detection~\cite{lin2026agentraft}; AgentSCOPE models privacy risks through workflow information flows~\cite{ngong2026agentscope}; and Agent-BOM proposes a unified graph representation for agent security auditing~\cite{li2026agentbom}.
These approaches show the need for agent aware analysis, but they target properties different from IALs, such as workflow topology, policy satisfaction, tool effects, sandboxing, privacy leakage, capability bindings, or runtime audit evidence. LLM assisted static analysis systems, including IRIS, QLCoder, and Argus, further show that LLMs can help infer specifications, synthesize static queries, or orchestrate vulnerability analysis~\cite{li2025iris,wang2026qlcoder,liang2026argus}. However, existing work does not specifically check whether an agentic feedback path can repeatedly trigger costly or state changing actions without an effective bound. Our work addresses this gap by combining framework aware feedback path recovery with bound coverage analysis for IAL detection.

\begin{figure*}[t]
    \centering
    \includegraphics[width=0.98\linewidth]{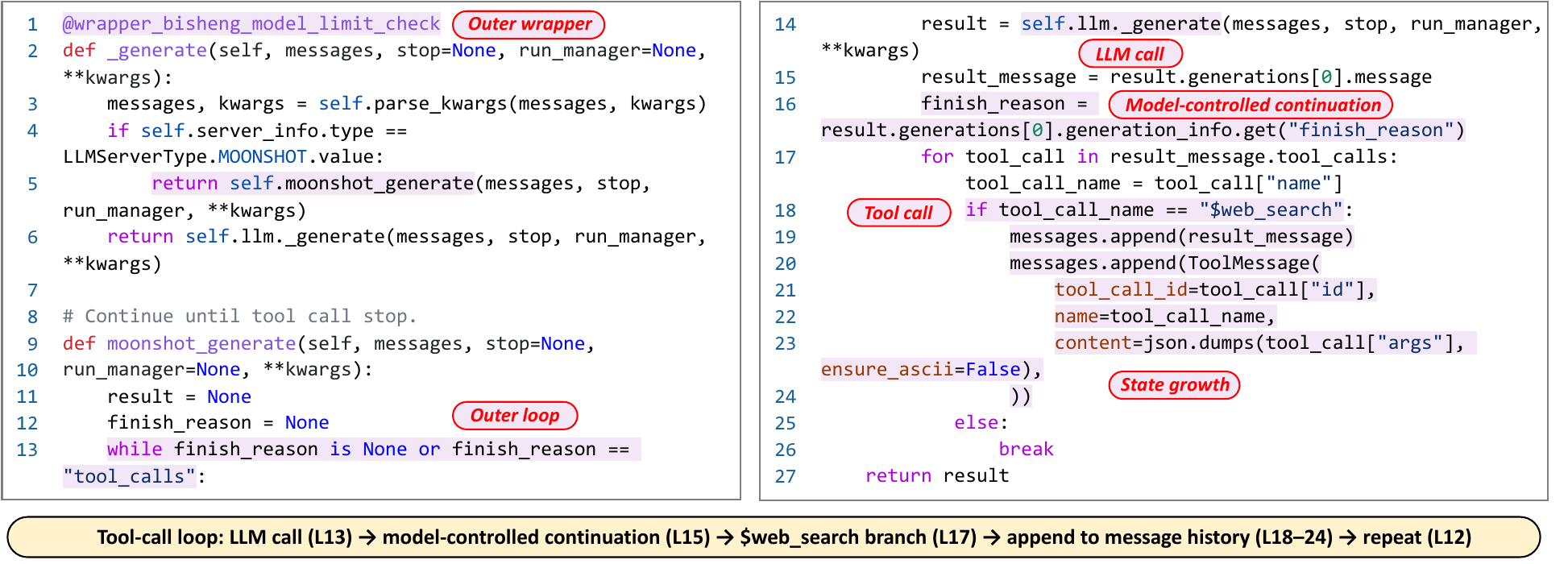}
    \caption{A motivating example of an IAL failure in \texttt{dataelement/bisheng}.}
    \label{fig:motivating}
\end{figure*}

\section{Problem Statement}
\label{sec:statement}

\begin{table*}[t]
\centering
\scriptsize
\caption{Representative interfaces for common agent loop behavior.}
\label{tab:challenge1}
\setlength{\tabcolsep}{5pt}
\resizebox{1\linewidth}{!}{
\begin{tabular}{c|l|l|c}
\hline
\textbf{Program Role} &
\textbf{Motivating Example (\autoref{fig:motivating})} &
\textbf{Other Framework Forms} &
\textbf{IAL Signal} \\
\hline

Model call &
\texttt{self.llm.\_generate(...)} &
\texttt{AgentExecutor.invoke(...)}; \texttt{Runner.run(...)} &
Costly invocation \\

Continuation control &
\texttt{finish\_reason == "tool\_calls"} &
\texttt{tools\_condition}; \texttt{termination\_condition} &
Model output controls loop \\

Tool dispatch &
\texttt{result\_message.tool\_calls}; \texttt{\$web\_search} &
\texttt{ToolNode(...)}; \texttt{@tool}; \texttt{tools=[...]} &
Feedback through tool call \\

State update &
\texttt{messages.append(...)}; \texttt{ToolMessage(...)} &
\texttt{chat\_history}; \texttt{memory}; \texttt{workflow state} &
Output reused as input \\

Bound &
No \texttt{max\_tool\_calls} or timeout &
\texttt{max\_iterations}; \texttt{max\_turns}; \texttt{recursion\_limit} &
Missing bound coverage \\

\hline
\end{tabular}}
\end{table*}

\subsection{Definition of Infinite Agentic Loop Failures}
We define an \textbf{\emph{Infinite Agentic Loop} (IAL)} failure as a structural execution failure where an agentic feedback path repeatedly triggers costly or state-growing actions without an effective stopping bound. An IAL is not merely a loop; it arises when model outputs, tool results, external observations, or delegation decisions can keep the path active, while no strong bound covers it.

\subsection{Threat Model}
\label{sec:threat-model}

\noindent \textbf{Assumptions.}
We consider deployed LLM agents that interact with users, call LLMs and tools, access external services, update state, and may perform side-effecting actions such as file writes, database updates, code execution, or ticket creation. We assume the agent code, framework configuration, and tool definitions are fixed before deployment, and that the underlying frameworks and trusted tool implementations are not malicious. However, execution safeguards such as iteration limits, retry caps, timeouts, recursion limits, human approval checks, or policy gates may be missing, ineffective, disabled, or only partially applied to the repeated feedback path.

\noindent \textbf{Attacker Capabilities.}
An attacker or untrusted user can interact with the agent through normal inputs, such as prompts, task requests, documents, URLs, issue reports, tickets, workflow parameters, or API calls. The attacker may craft inputs that influence model outputs, tool arguments, retrieved content, external observations, or error conditions, causing the agent to repeat tool calls, retries, polling, sub-agent invocations, or workflow transitions. The attacker cannot modify the source code, framework configuration, deployment environment, or trusted tool implementation.

\noindent \textbf{Scope.}
We focus on IAL failures that are structurally visible in source code. We include both core agent feedback loops and toolchain loops that can block, reenter, or amplify agent execution. We exclude high-volume network DoS, infrastructure-level resource exhaustion, prompt sponge attacks, malicious framework modifications, and purely semantic no-progress behavior that cannot be inferred statically.

\begin{figure*}[t]
    \centering
    \includegraphics[width=0.9\linewidth]{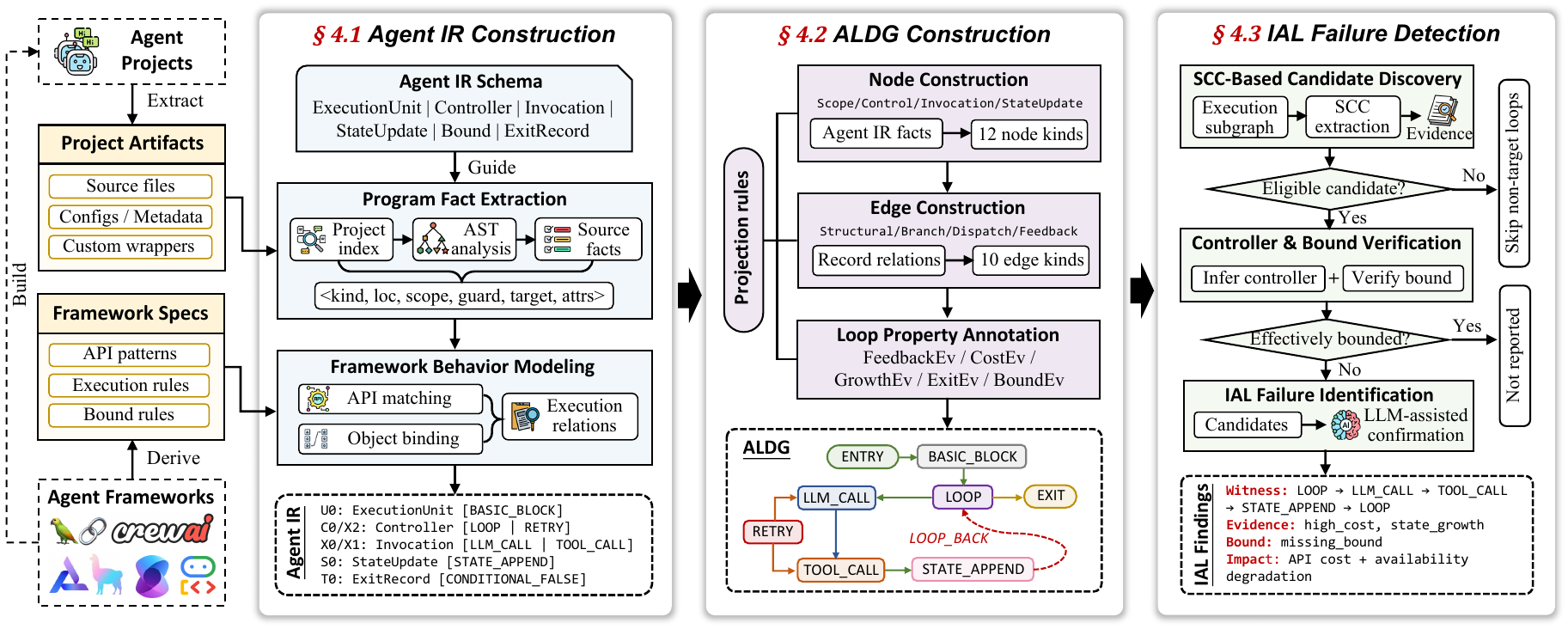}
    \caption{Overview of \toolname{} pipeline.}
    \label{fig:overview}
\end{figure*}

\subsection{Motivating Example}

\autoref{fig:motivating} shows an IAL failure example. The code implements a custom model wrapper compatible with the LangChain \texttt{BaseChatModel} interface and dispatches Moonshot requests to \texttt{moonshot\_generate}. The repeated behavior is therefore hidden inside a model wrapper rather than exposed as a standalone agent graph.
In \texttt{moonshot\_generate}, the outer loop continues while \texttt{finish\_reason} is \texttt{None} or \texttt{``tool\_calls''}. Each iteration invokes the model through \texttt{self.llm.\_generate(...)} and reads \texttt{finish\_reason} from the returned message. When the model returns a \texttt{\$web\_search} tool call, the code appends both the model message and a \texttt{ToolMessage} to \texttt{messages}. The next iteration sends the enlarged message history back to the model. This forms a feedback path from model output to state update and then back to the next model call.
Although the loop may exit when the model stops producing \texttt{``tool\_calls''}, the exit is controlled by model output and is not a deterministic bound. The implementation does not enforce a local \texttt{max\_tool\_calls}, \texttt{max\_iterations}, or timeout over this feedback path. The inner \texttt{break} only exits the tool iteration, not the outer loop. This agent may repeatedly invoke the model while growing the message history, increasing latency, token usage, API cost, and worker occupation.

\subsection{Challenges in Detecting IAL Failures}

IAL detection requires reasoning beyond syntactic loops. In real agent applications, repeated execution may be encoded through framework APIs, indirect runtime dispatch, state updates, and configuration values. We summarize three challenges.

\subsubsection{Agent Behavior Encoded by Frameworks} 
Agent behavior is often expressed through framework interfaces rather than direct calls. As shown in \autoref{tab:challenge1}, model execution, continuation control, tool dispatch, state update, and stopping mechanisms may appear through different APIs across agent frameworks. For example, similar behavior may be encoded by \texttt{AgentExecutor.invoke(...)}, \texttt{ToolNode(...)}, \texttt{tools\_condition}, \texttt{Runner.run(...)}, delegation APIs, or group chat termination logic. A detector must map these framework forms to common execution concepts, instead of relying on one API name or one syntactic pattern. 

\subsubsection{Feedback Path Reconstruction} 
An IAL failure is usually not explained by one statement. In \autoref{fig:motivating}, the relevant path connects the model call, continuation guard, tool branch, message update, and repeated execution. A static analysis must preserve control, call, and state relations across wrapper code and framework logic, while retaining guard dependencies such as model status values, tool call fields, and routing predicates. 

\subsubsection{Bound Coverage Reasoning} 
Loops are common in agent applications, but they become problematic when a repeated path can reach costly or state-growing operations without an effective bound. As illustrated by the missing bound in \autoref{fig:motivating} and the representative bound forms in \autoref{tab:challenge1}, detection must check whether a bound constrains the controller or a scope that covers the feedback path, rather than merely checking whether a limit or exit condition appears near the loop.

\section{Design of \toolname{}}
\label{sec:methodology}

In this section, we present \toolname{}. To address the challenges above, \toolname{} first builds a framework-independent \emph{Agent IR}, then constructs an \emph{Agentic Loop Dependence Graph} (ALDG), and finally detects feedback paths that may repeatedly trigger costly or state-growing actions without an effective stopping mechanism. \autoref{fig:overview} summarizes the pipeline.

\subsection{Agent IR Construction}
\label{sec:agent-ir-construction}

\toolname{} first builds Agent IR, an intermediate representation independent of any specific agent framework. 

\subsubsection{Agent IR Schema}
\label{sec:agent-ir-schema}

Agent IR represents an agent project as typed facts and local relations. It abstracts over different encodings of agent execution in source code and framework APIs: repeated execution may appear as an explicit loop, a workflow transition, a framework handoff, or a tool dispatch. As summarized in \autoref{lst:agent-ir-schema}, the fact sets capture the main program elements, while the relation facts record how these elements are connected through source code and framework APIs. \texttt{ExecutionUnit} facts define execution scopes; \texttt{Controller} facts capture loops, routers, retry logic, and termination predicates; \texttt{Invocation} facts represent model calls, tool calls, agent runs, workflow runs, and subprocesses; and \texttt{StateUpdate} facts record state changes that may persist across iterations. \texttt{Bound} and \texttt{ExitRecord} facts capture potential stopping mechanisms. Local relations connect these facts through ownership, calls, updates, guard dependencies, exits, bounds, workflow transitions, tool dispatches, and aliases.

\begin{lstlisting}[
  style=AgentIRSchema,
  caption={Core Schema in Agent IR.},
  label={lst:agent-ir-schema}
]
AgentIR:
  facts:
    execution_units: set[ExecutionUnit]
    controllers: set[Controller]
    invocations: set[Invocation]
    state_updates: set[StateUpdate]
    bounds: set[Bound]
    exits: set[ExitRecord]

  common fields:
    id, kind, location, attrs

  relations:
    owns(unit, fact)
    calls(invocation, callee)
    updates(state_update, target)
    guards(controller, variable)
    exits(exit_record, controller)
    constrains(bound, target)
    transitions(source, target)
    dispatches(invocation, target)
    aliases(name, target)
\end{lstlisting}

\subsubsection{Program Fact Extraction}
\label{sec:source-program-analysis}

For each agent project, \toolname{} builds a project index and parses Python files into ASTs. The index records modules, imports, framework usage, decorators, local factories, custom runtime classes, and project-defined agent callables. It also performs lightweight name and attribute resolution for import aliases, local assignments, object fields, and factory returns. This resolution helps identify candidate call targets and state objects without whole-program points-to analysis.
The AST pass extracts local control and data facts, including loops, recursive calls, call expressions, state updates, conditional exits, and \texttt{try}/\texttt{except} retry paths. Each source fact has the form
\(
\langle \mathrm{kind}, \mathrm{loc}, \mathrm{scope}, \mathrm{guard}, \mathrm{target}, \mathrm{attrs} \rangle ,
\)
where \(\mathrm{scope}\) records the enclosing function, class, agent, or workflow, \(\mathrm{guard}\) stores the relevant condition, and \(\mathrm{target}\) records the resolved callee, updated object, bounded variable, or exception type. The extracted facts are translated into Agent IR: loops become \texttt{Controller} facts, runtime calls become \texttt{Invocation} facts, loop-carried updates become \texttt{StateUpdate} facts, caps and budgets become \texttt{Bound} facts, and local exits become \texttt{ExitRecord} facts. In \autoref{fig:motivating}, \texttt{self.llm.\_generate(...)} becomes an \texttt{LLM\_CALL}, \texttt{messages.append(...)} becomes a \texttt{STATE\_APPEND}, and \texttt{finish\_reason} and \texttt{tool\_calls} become guard dependencies of the surrounding controller.

\subsubsection{Framework Behavior Modeling}
\label{sec:framework-behavior-modeling}

\begin{lstlisting}[
  style=AgentIRExample,
  caption={Excerpt of Agent IR facts for the motivating example.},
  label={lst:motivating-agent-ir},
  float=t
]
ExecutionUnit:
  U0: FUNCTION; label=moonshot_generate;
      framework=custom_python

Controller:
  C0: LOOP; owner=U0;
      condition=finish_reason is None
          or finish_reason == "tool_calls";
      guard_deps={finish_reason, 
          result_message.tool_calls};
      guard_source=model_or_tool_output

Invocation:
  X0: LLM_CALL; owner=U0; callee=self.llm._generate;
      attrs={high_cost=true}

StateUpdate:
  S0: STATE_APPEND; owner=U0; target=messages;
      attrs={state_growth=true}

ExitRecord:
  T0: CONDITIONAL_FALSE; owner=C0; target=return;
      condition=finish_reason != "tool_calls"
\end{lstlisting}

Explicit source constructs do not expose all execution behavior in agent programs. A tool may be registered in one place and invoked later by a dispatcher; a workflow edge may execute a node without a direct call; and a handoff or delegation may transfer control to another agent. Using the resolution results from the previous step, \toolname{} models such framework behavior as Agent IR facts and relations.
The models handle construction, invocation, and configuration. Construction models identify framework objects and bindings, including agents, tools, workflows, graph nodes, and runtime scopes. Invocation models derive implicit execution relations, such as tool dispatch, workflow transitions, handoffs, delegation, and agent reentry through \texttt{Agent.as\_tool(...)}. Configuration models attach limits such as \texttt{max\_turns}, \texttt{max\_iterations}, \texttt{max\_retry}, and \texttt{recursion\_limit} to the corresponding runtime scope or repeated path.
For example, LangGraph nodes and conditional edges become workflow scopes, routing controllers, and transition relations; OpenAI Agents SDK \texttt{Runner.run(...)}, handoffs, and \texttt{Agent.as\_tool(...)} become invocations and reentry relations; and CrewAI delegation becomes agent execution with a delegation relation. If a target cannot be resolved statically, \toolname{} preserves it as an unresolved attribute rather than creating a precise transfer relation.
\autoref{lst:motivating-agent-ir} shows an excerpt of the Agent IR facts for \autoref{fig:motivating}. The excerpt includes the enclosing execution unit, loop controller, model call, state update, and local exit condition.

\subsection{Agentic Loop Dependence Graph (ALDG) Construction}
\label{sec:aldg-construction}

Given Agent IR, \toolname{} constructs an Agentic Loop Dependence Graph (ALDG), a directed attributed graph for reasoning about feedback paths in agent execution. ALDG keeps the facts needed to connect controllers, runtime actions, state updates, exits, and bounds. \autoref{fig:aldg-rules} summarizes the construction rules.

\begin{figure}[t]
\centering
\fbox{
\begin{minipage}{0.99\linewidth}
\footnotesize
\textbf{Notations.}
\begin{itemize}[leftmargin=12pt]
    \setlength{\itemsep}{1pt}
    \setlength{\parskip}{0pt}
    \item \(G_A=(V_A,E_A,\alpha)\): ALDG with vertices, edges, and attributes.
    \item \(r\): an Agent IR fact; \(v_r\): the ALDG vertex created for \(r\).
    \item \(\mathcal{C}\), \(\mathcal{I}_{\mathrm{cost}}\), \(\mathcal{S}_{\mathrm{grow}}\), and \(\mathcal{U}_{\mathrm{scope}}\): controllers, costly invocations, growing state updates, and scope facts.
    \item \(\mathcal{T}_{V}\): ALDG node kinds; \(\mathcal{K}_{E}\): ALDG edge kinds.
    \item \(r_i \xrightarrow{\kappa,\Delta} r_j\): Agent IR relations imply an ALDG edge of kind \(\kappa\), with attributes \(\Delta\).
    \item \(\mathrm{body}(c)\), \(\mathrm{cycle}(c)\): facts and edges in the body and feedback component of controller \(c\).
\end{itemize}

\vspace{3pt}

\textbf{Construction Rules.}
\setlength{\abovedisplayskip}{2.5pt}
\setlength{\belowdisplayskip}{2.5pt}
\begin{enumerate}[leftmargin=12pt]
\setlength{\itemsep}{2pt}
\setlength{\parskip}{0pt}

\item \textit{\textbf{Node Construction.}}
\[
\begin{aligned}
r &\in
\mathcal{C}\cup\mathcal{I}_{\mathrm{cost}}
\cup\mathcal{S}_{\mathrm{grow}}
\cup\mathcal{U}_{\mathrm{scope}} \\
&\quad\Rightarrow\quad
v_r \in V_A,\;
\mathrm{kind}(v_r)\in\mathcal{T}_{V},\;
\alpha(v_r)\leftarrow
\mathrm{fields}(r)\cup\mathrm{attrs}(r).
\end{aligned}
\]

\item \textit{\textbf{Edge Construction.}}
\[
r_i \xrightarrow{\kappa,\Delta} r_j,\;
\kappa\in\mathcal{K}_{E}
\quad\Rightarrow\quad
(v_{r_i},v_{r_j},\kappa)\in E_A,\;
\alpha(v_{r_i},v_{r_j})\leftarrow\Delta.
\]

\item \textit{\textbf{Loop Summary Construction.}}
\[
\begin{aligned}
\mathrm{Feedback}(c)
&= \{\mathrm{kind}(e)\mid e\in\mathrm{cycle}(c),\;
   e.\mathrm{feedback}\} \\ &
   \cup\{\mathrm{guard\_source}(c)\},\\
\mathrm{Cost}(c)
&= \{\mathrm{kind}(i)\mid i\in\mathrm{body}(c),\;
   i.\mathrm{high\_cost}\},\\
\mathrm{Growth}(c)
&= \exists s\in\mathrm{body}(c):s.\mathrm{state\_growth},\\
\mathrm{Exit}(c)
&= \{x\mid x.\mathrm{owner}=c
   \vee x.\mathrm{target}=c\},\\
\mathrm{Bound}(c)
&= \{b\mid b \text{ constrains } c
   \text{ or its owner scope}\}.
\end{aligned}
\]

\end{enumerate}
\end{minipage}
}
\caption{Construction rules from Agent IR to ALDG.}
\label{fig:aldg-rules}
\end{figure}

\subsubsection{ALDG Node Construction}
\label{sec:aldg-node-construction}

\begin{figure}[t]
\centering
\includegraphics[width=0.99\linewidth]{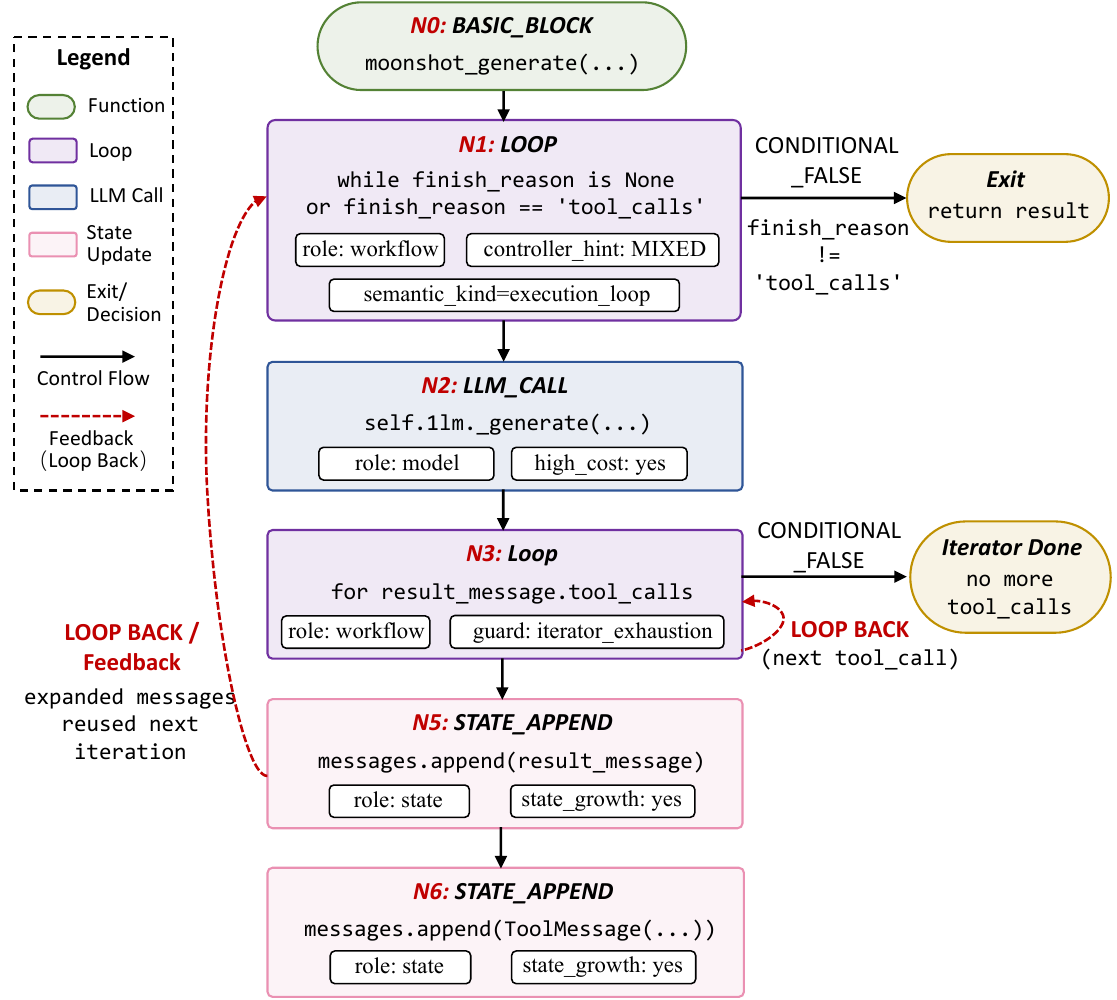}
\caption{ALDG for the motivating example in \autoref{fig:motivating}.}
\label{fig:aldg-motivating}
\end{figure}

\toolname{} derives ALDG nodes from Agent IR facts that are relevant to loop behavior. As defined in \autoref{fig:aldg-rules}, the retained fact sets are controllers \(\mathcal{C}\), costly invocations \(\mathcal{I}_{\mathrm{cost}}\), growing state updates \(\mathcal{S}_{\mathrm{grow}}\), and execution scopes \(\mathcal{U}_{\mathrm{scope}}\). These facts are mapped to 12 node kinds in \(\mathcal{T}_{V}\), covering scope, controller, invocation, and state nodes. They correspond to the main questions used in later analysis: where repetition is controlled, which actions may consume resources, which state may grow across iterations, and which scope owns the behavior.
For each retained fact \(r\), \toolname{} creates a node \(v_r\), assigns a node kind from \(\mathcal{T}_{V}\), and copies the source location and analysis attributes of \(r\) to the node. Guard variables, configuration values, and aliases are kept as node attributes when they help interpret the corresponding node. \autoref{fig:aldg-motivating} shows the ALDG for the motivating example: \texttt{self.llm.\_generate(...)} and \texttt{messages.append(...)} become \texttt{LLM\_CALL} and \texttt{STATE\_APPEND} nodes, while \texttt{finish\_reason} and \texttt{tool\_calls} are stored as guard attributes of the surrounding \texttt{LOOP} node.

\subsubsection{ALDG Edge Construction}
\label{sec:aldg-edge-construction}

\toolname{} derives typed ALDG edges from Agent IR relations among retained nodes. Since auxiliary IR elements such as guards, aliases, configuration entries, and intermediate statements may not appear as nodes, edge construction preserves reachability through these elements and assigns each derived relation an edge kind in \(\mathcal{K}_{E}\).
We define 10 edge kinds in \(\mathcal{K}_{E}\), grouped as execution, exit, framework, and feedback edges. For two retained Agent IR facts \(r_i\) and \(r_j\), a derived relation
\(r_i \xrightarrow{\kappa,\Delta} r_j\), where \(\kappa\in\mathcal{K}_{E}\), yields an ALDG edge \((v_{r_i},v_{r_j},\kappa)\). The attributes \(\Delta\) record source locations, guard conditions, resolved targets, and framework bindings.
Source structure gives \texttt{CONTROL\_FLOW} edges, while resolved calls give \texttt{CALL} edges. Framework relations give \texttt{WORKFLOW\_TRANSITION} and \texttt{TOOL\_DISPATCH} edges after matching tool registrations, graph nodes, routing functions, and runtime scopes. Loop back, recursion, agent reentry, and exception retry become feedback edges, and exits become \texttt{CONDITIONAL\_TRUE} or \texttt{CONDITIONAL\_FALSE} edges.
In \autoref{fig:aldg-motivating}, the wrapper scope reaches the loop, model call, and state update through execution edges. The outer loop has a \texttt{LOOP\_BACK} edge and a \texttt{CONDITIONAL\_FALSE} edge to the exit node. This exit is retained but not counted as a bound because it depends on model output.

\subsubsection{Loop Property Annotation} 
\label{sec:loop-property-annotation}

For each controller node \(c\), \toolname{} annotates the node with loop properties computed from its reachable body and incident ALDG edges. These properties record whether the controlled region reaches costly actions, updates state carried across iterations, exposes local exits, or has candidate bounds in a matching scope. They are stored as controller attributes, including \texttt{body\_cost\_kinds}, \texttt{state\_growth}, \texttt{exit\_kinds}, \texttt{guard\_source}, and \texttt{bound\_sources}. Edge attributes preserve relation information such as edge role, guard condition, feedback kind, and source locations.
These loop properties are not final IAL decisions; they provide graph-level inputs to the failure checker. In the ALDG example shown in \autoref{fig:aldg-motivating}, the outer loop has a visible \texttt{CONDITIONAL\_FALSE} exit, but the exit depends on the model-produced \texttt{finish\_reason}. \toolname{} therefore records it as a model-dependent exit rather than a deterministic bound. Candidate bounds are recorded in the same way, while effective bound coverage is checked during failure detection.

\subsection{Infinite Agentic Loop (IAL) Failure Detection}
\label{sec:failure-detection}

\toolname{} detects IAL failures over the ALDG by checking whether an execution feedback path can repeatedly reach agentic actions, costly invocations, or growing state without an effective stopping mechanism, as shown in \autoref{alg:ial-detection}.

\begin{algorithm}[t]
\caption{IAL failure detection over ALDG}
\label{alg:ial-detection}
\footnotesize
\begin{algorithmic}[1]
\Require ALDG \(G_A=(V_A,E_A,\alpha)\), cycle edge kinds \(\mathcal{K}_{\mathrm{cycle}}\)
\Ensure IAL findings \(\mathcal{F}\)

\State \(\mathcal{F}\gets \emptyset\)
\State \(G_{\mathrm{cycle}}\gets (V_A,\{e\in E_A\mid \mathrm{kind}(e)\in\mathcal{K}_{\mathrm{cycle}}\})\)
\State \(\mathcal{L}\gets \{L\in \mathrm{SCC}(G_{\mathrm{cycle}})\mid |L|>1 \vee \mathrm{SelfLoop}(L)\}\)

\ForAll{\(L\in\mathcal{L}\)}
    \State \(\Pi_L\gets \Call{CollectProperties}{L,G_A}\)
    \State \(\Theta_L\gets \Call{CollectTopology}{L,G_A}\)

    \If{\(\neg\Call{EligibleCandidate}{L,\Pi_L,\Theta_L}\)}
        \State \textbf{continue}
    \EndIf

    \State \(c_L\gets \Call{SelectController}{L,\Pi_L}\)
    \State \(\beta_L\gets \Call{CheckBoundCoverage}{L,c_L,\Pi_L,\Theta_L}\)

    \If{\(\Call{IsCovered}{\beta_L}\)}
        \State \textbf{continue}
    \EndIf

    \State \(F_L\gets \Call{BuildFinding}{L,c_L,\beta_L,\Pi_L,\Theta_L}\)

    \If{\(\Call{Reportable}{F_L}\)}
        \State \(\mathcal{F}\gets \mathcal{F}\cup\{F_L\}\)
    \EndIf
\EndFor

\State \Return \(\mathcal{F}\)
\end{algorithmic}
\end{algorithm}

\subsubsection{SCC-Based Candidate Discovery}
\label{sec:cycle-discovery}

\toolname{} first derives a cycle-relevant subgraph from the ALDG. The subgraph keeps edges that may participate in repeated execution, including control flow, calls, framework transitions, tool dispatch, and feedback edges. Exit edges and bound attributes are excluded from SCC construction, because they describe possible stopping conditions rather than repeated execution itself. \toolname{} then computes strongly connected components and retains nontrivial SCCs, including singleton SCCs with explicit self-loop edges.
Each retained SCC is treated as a candidate feedback region. \toolname{} computes a property set \(\Pi_L\) and a topology set \(\Theta_L\) for the candidate. \(\Pi_L\) records entry reachability, feedback edges, costly invocations, state growth, local exits, and guard dependencies. \(\Theta_L\) records workflow structure, such as routing cycles, missing exit transitions, dynamic dispatch, and unreachable exits. A candidate is kept only if it is reachable from an agent entry point and contains a feedback path that reaches agentic actions, costly invocations, or growing state. Cycles that correspond to bounded iteration patterns, stream consumers, parsers, pagination loops, lifecycle loops, or test scaffolding are filtered before bound verification.

\subsubsection{Controller and Bound Verification}
\label{sec:controller-bound-analysis}

For each candidate \(L\), \toolname{} selects the continuation controller \(c_L\) using controller nodes, guard dependencies, exit predicates, and feedback edge kinds. The controller is classified as deterministic, model controlled, tool controlled, external state controlled, exception controlled, or mixed. This classification matters because a visible exit does not necessarily stop the feedback path when continuation depends on model output, tool results, exceptions, or remote state.
\toolname{} then checks whether an effective bound covers the candidate feedback path. As summarized in \autoref{lst:bound-status}, each candidate is assigned a bound status. A bound records its kind, value, source, strength, and target, but it is effective only when it constrains the repeated path itself. The check verifies whether a bound applies to the controller, its runtime scope, or the feedback path it controls. For example, an inner turn cap on a nested agent call does not cover an outer evaluator feedback cycle unless it dominates the outer feedback path.

\begin{lstlisting}[
  style=AgentIRExample,
  caption={Bound coverage status for candidate feedback paths.},
  label={lst:bound-status}
]
BoundStatus:
  Covered:
    verified_bound | framework_default_bound
    | config_dependent_bound
  UncoveredOrWeak:
    missing_bound | weak_bound | disabled_bound
    | ineffective_bound | bypassed_bound
\end{lstlisting}
Lines~12--14 skip candidates whose feedback paths are covered by an effective bound. The remaining candidates are treated as unbounded or ineffectively bounded feedback paths.

\subsubsection{IAL Failure Identification}
\label{sec:finding-classification}

\toolname{} reports a candidate as an IAL failure only when it satisfies three conditions: it contains an agentic feedback path, the path reaches costly or state-growing operations, and the repeated execution is not covered by an effective bound. Confidence increases with model-controlled continuation, repeated tool or agent execution, state growth, and the absence of a reachable deterministic exit. It decreases for non-target cycles such as stream consumers, lifecycle loops, pagination or parser loops, context-pruning loops, non-production polling, and test or example code.
For ambiguous candidates, \toolname{} applies an optional LLM-assisted pruning pass. For each candidate \(L\), it constructs a bounded slice \(S_L\) containing the feedback witness, controller condition, guard definitions, candidate bounds, relevant call chain, and source snippets. The LLM acts only as a negative filter over \(S_L\): it may suggest pruning predicates such as \texttt{strong\_finite\_bound}, \texttt{non\_agentic\_loop}, \texttt{test\_only\_code}, or \texttt{deterministic\_exit}. A predicate is accepted only if it is supported by the slice and does not contradict the static properties of \(L\). 
\section{Evaluation}
\label{sec:evaluation}

In this section, we evaluate \toolname{} with the following research questions~(RQs):

\noindent\hangindent=2.5em\hangafter=1
\textbf{RQ1 [Real-world Findings]}
What IAL failures does \toolname{} identify in real-world LLM agent repositories, and what are their characteristics?

\noindent\hangindent=2.5em\hangafter=1
\textbf{RQ2 [Effectiveness]}
Which analysis components are necessary for finding IAL failures, and how does \toolname{} compare with LLM and coding-agent baselines?

\noindent\hangindent=2.5em\hangafter=1
\textbf{RQ3 [Stability]}
How stable are the static candidate generation stage and the LLM-assisted pruning stage across repeated runs and model choices?

\subsection{Experimental Setup}

\noindent\textbf{Implementation.}
We implement \toolname{} in Python. It constructs Agent IR, builds an ALDG, and checks feedback cycles for reachability, continuation control, bound coverage, resource consumption, and state carried across repeated executions. The LLM-assisted pruning stage uses GPT-5.5 as the default model. \toolname{} currently supports the analysis of downstream agent projects built with eight frameworks: LangChain, LangGraph, CrewAI, AutoGen, LlamaIndex, the OpenAI Agents SDK, Google ADK, and Semantic Kernel.

\noindent\textbf{Running environment.}
All experiments are conducted on a single Ubuntu 24.04.3 LTS server with two AMD EPYC 9554 processors, 256 logical CPU cores, and six NVIDIA A100 GPUs with 80\,GB memory each. 

\noindent\textbf{Dataset.}
Our evaluation uses a corpus of 6,549 Python LLM agent repositories with at least one GitHub star, collected from GitHub metadata and cloned snapshots. We identify candidates by searching dependency declarations, imports, API uses, and orchestration patterns for eight agent frameworks, and then filter them using repository metadata and source code evidence. Each retained repository must be a downstream application or product, rather than a framework implementation, tutorial, or isolated example, and must contain concrete agent logic such as agent or workflow construction, runtime invocation, tool registration, or orchestration. The corpus contains 246,748 Python files and 33.41M lines of Python code.

\noindent\textbf{Manual review protocol.}
Manual review is used for confirming real-world findings in RQ1 and identifying missed cases in RQ2. The first two authors independently inspect each case and label it as a confirmed IAL failure, a false positive, or a missed case of \toolname{}. A case is confirmed as an IAL failure if it contains a repeated feedback path involving model, tool, agent, or workflow execution, its continuation depends on runtime outputs, and no strong bound covers the repeated path. Disagreements are resolved through discussion, with another author consulted when needed.

\begin{table}[t]
\centering
\caption{Distribution of 68 confirmed IAL failures.}
\label{tab:rq1-finding-distribution}
\resizebox{1\linewidth}{!}{
\begin{tabular}{lrrr}
\hline
\textbf{Category} & \textbf{\#Count} & \textbf{Ratio} & \textbf{\#Proj.} \\
\hline
\rowcolor{gray!8}
\multicolumn{4}{c}{\textbf{Framework}} \\ \hline
LangGraph & 23 & 33.8\% & 16 \\
AutoGen AgentChat & 22 & 32.4\% & 15 \\
LlamaIndex & 6 & 8.8\% & 2 \\
LangChain AgentExecutor & 5 & 7.4\% & 3 \\
CrewAI & 4 & 5.9\% & 4 \\
OpenAI Agents SDK & 4 & 5.9\% & 3 \\
Google ADK & 3 & 4.4\% & 3 \\
Semantic Kernel & 1 & 1.5\% & 1 \\
\hline
\rowcolor{gray!8}
\multicolumn{4}{c}{\textbf{Failure Pattern}} \\ \hline
Retry feedback without bound & 17 & 25.0\% & 10 \\
Tool-call iteration without bound & 16 & 23.5\% & 11 \\
Multi-agent chat without turn bound & 14 & 20.6\% & 13 \\
Workflow loop without effective bound & 9 & 13.2\% & 8 \\
Message reentry without bound & 7 & 10.3\% & 2 \\
Runner, delegation, or evaluator feedback & 5 & 7.4\% & 4 \\
\hline
\rowcolor{gray!8}
\multicolumn{4}{c}{\textbf{Impact}} \\ \hline
API cost exhaustion & 65 & 95.6\% & 44 \\
Model denial of service & 65 & 95.6\% & 44 \\
Context window exhaustion & 19 & 27.9\% & 14 \\
External tool rate-limit exhaustion & 5 & 7.4\% & 5 \\
\hline
\rowcolor{gray!8}
\multicolumn{4}{c}{\textbf{Root Cause}} \\ \hline
Missing strong bound & 68 & 100.0\% & 47 \\
Tool-controlled retry & 28 & 41.2\% & 18 \\
Model-controlled termination & 26 & 38.2\% & 19 \\
Missing exit & 23 & 33.8\% & 17 \\
Workflow cycle without verified bound & 21 & 30.9\% & 15 \\
State growth amplifier & 19 & 27.9\% & 14 \\
Agent tool reentry & 17 & 25.0\% & 11 \\
\hline
\end{tabular}}
\end{table}

\subsection{RQ1: Real-world Findings}
\label{sec:rq1-detection}

\subsubsection{Overall Result}
On the real-world corpus of 6,549 LLM agent projects, \toolname{} reports 74 potential findings. Manual review confirms 68 IAL failures and 6 false positives, yielding an end-to-end precision of 91.9\%. During independent labeling, the first two authors agreed on 94.6\% of the 74 potential findings; the remaining cases were resolved through discussion.
These failures affect 47 agent projects and cover all eight modeled framework families. 
As shown in \autoref{tab:rq1-finding-distribution}, LangGraph and AutoGen contribute 45 of the 68 confirmed findings (66.2\%) across 31 projects. Both frameworks encode feedback through APIs rather than visible loop syntax. For LangGraph, \toolname{} models APIs such as \texttt{add\_edge}, \texttt{add\_conditional\_edges}, and \texttt{tools\_condition} to capture LLM--tool feedback paths. For AutoGen, it models \texttt{GroupChat}, \texttt{initiate\_chat}, and termination predicates such as \texttt{max\_turns} and \texttt{is\_termination\_msg} to capture agent-to-agent feedback paths.
The remaining 23 findings span the other six frameworks, suggesting that IAL failures are not specific to one framework or loop idiom.
Retry feedback without bounds, tool-call iteration without bounds, and multi-agent chat without turn bounds account for 47 findings (69.1\%). These failures often arise when parser errors, validator failures, repeated tool requests, or generated agent messages redirect execution to model, tool, or agent actions. 
The dominant impacts are API cost exhaustion and model denial of service, each appearing in 95.6\% findings. Another 19 findings may exhaust the context window due to message or workflow state growth. All 68 failures share the same root issue: the repeated path is not covered by a strong bound. This gap commonly appears in tool-controlled retries, model-dependent termination, and workflow or agent reentry without verified limits.

\subsubsection{Case Study}
\begin{figure}[t]
    \centering
    \includegraphics[width=0.9\linewidth]{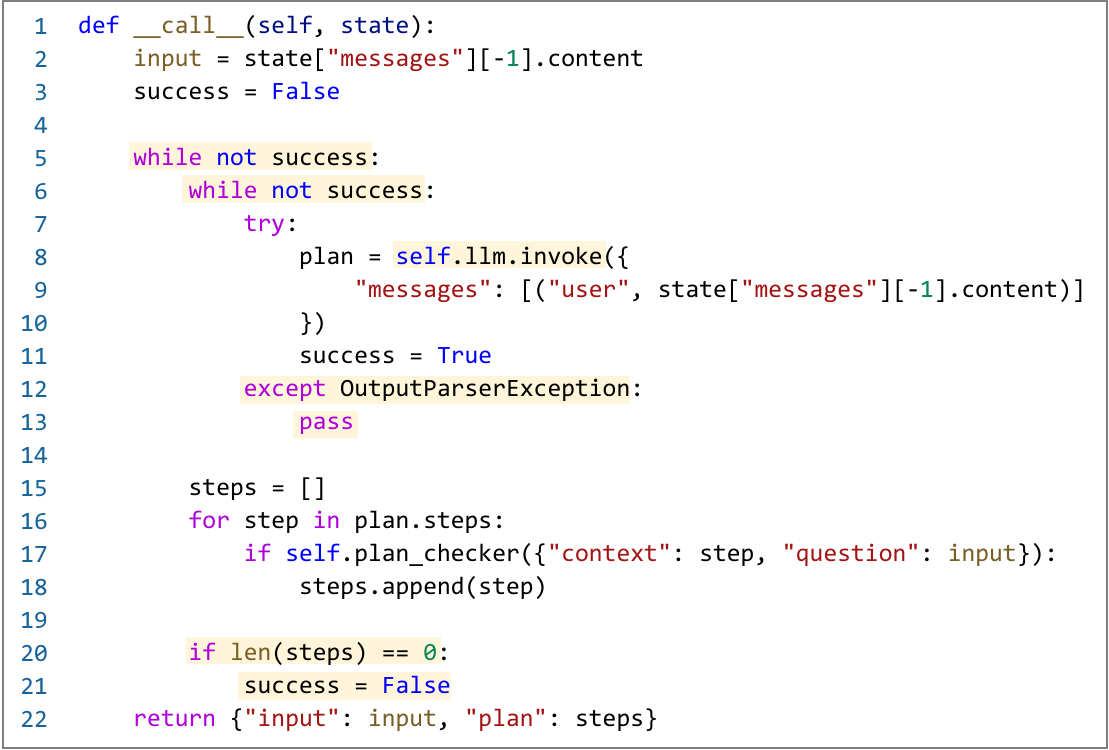}
    \caption{Retry loop missing bound.}
    \label{fig:case1}
\end{figure}

We further examine two confirmed findings to illustrate different IAL patterns. \autoref{fig:case1} shows a retry loop in \texttt{2456868764/LiteRAG}. The planner uses nested \texttt{while not success} loops to repeatedly request a plan from the LLM. The costly call appears at lines~8--10, where \texttt{self.llm.invoke(...)} is executed until parsing succeeds. However, parser failures are swallowed at lines~12--13, and rejected plans reset \texttt{success} to \texttt{False} at lines~20--21. This sends execution back to the same LLM call. Since no retry cap, timeout, or token budget covers this feedback path, malformed or repeatedly rejected model outputs can cause repeated model invocations, leading to API cost exhaustion and model service occupation.

\begin{figure}[htbp]
    \centering
    \includegraphics[width=0.9\linewidth]{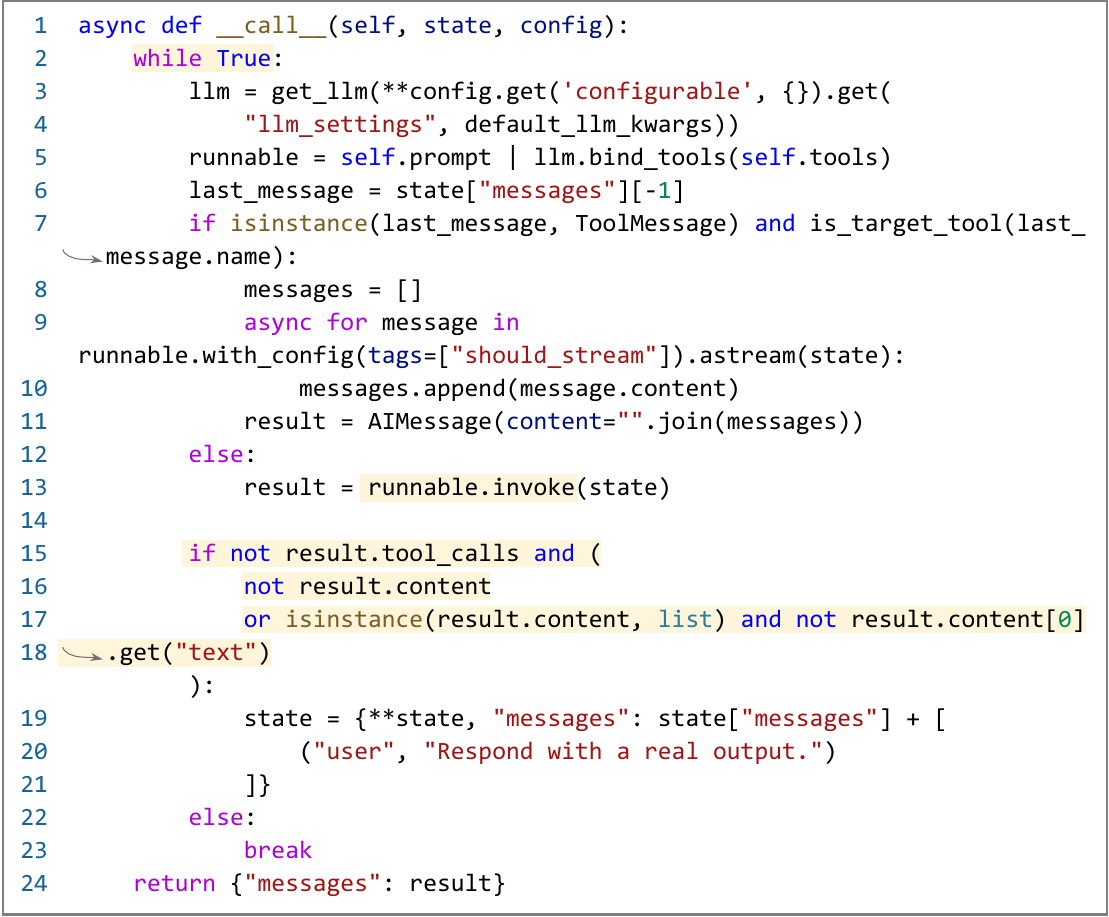}
    \caption{Tool call iteration missing bound.}
    \label{fig:case2}
\end{figure}

For the second case, \autoref{fig:case2} shows code snippet from \texttt{NVIDIA-AI-Blueprints/ai-virtual-assistant}, illustrating a tool call iteration failure. The assistant enters an unbounded \texttt{while True} loop at line~2, binds tools to the LLM at line~5, and invokes or streams the model at lines~9--13. The continuation check at lines~15--18 depends on whether the model returns tool calls or usable text. If the output is empty or malformed, the code appends a corrective prompt to \texttt{state["messages"]} at lines~19--21 and retries. This creates a feedback path from model output to message state growth and then back to another model call. Without a retry cap, tool call budget, timeout, or context size guard, the loop can amplify API cost, occupy model service capacity, and increase context window pressure.

\subsection{RQ2: Effectiveness}
\label{sec:rq2-effectiveness}

\subsubsection{Ablation Study}
Applying the full configuration of \toolname{} to the full corpus yields 340 static candidates across 264 projects. These candidates are then processed by LLM-assisted pruning and manual review, producing the end-to-end RQ1 result: 74 potential findings, 68 of which are true positives. We use these 264 projects as the evaluation subset for RQ2 and RQ3, as they cover the complete static candidate set generated by the full configuration and allow comparisons on the same set of relevant projects.

For the ablation study, we disable one component at a time. The four static analysis variants regenerate candidates after removing framework modeling, the agentic gate, bound coverage, or benign-loop filtering, and then apply the same LLM-assisted pruning stage. In contrast, \emph{w/o LLM Pruning} keeps the full static analysis stage and reports all static candidates directly.
The results in \autoref{tab:rq2-ablation} show that each component has a distinct role. Without \emph{LLM Pruning}, \toolname{} retains all 68 true positives but reports all 340 static candidates, producing 272 false positives. This confirms that the static stage favors recall and needs pruning to reduce manual review effort. \emph{Framework Modeling} improves precision by capturing framework specific feedback semantics; disabling it increases static candidates from 340 to 910 and alerts from 74 to 276, while reducing TP coverage from 68 to 61. The \emph{Agentic Gate} mainly controls cost: removing it causes the largest candidate increase, from 340 to 1,453, and raises token usage from 4.2K to 40.4K. \emph{Bound Coverage} affects both precision and recall, as removing it lowers TP coverage to 60 and increases false positives to 10. \emph{Benign Loop Filtering} suppresses safe loop contexts, reducing alerts from 103 to 74 and token usage from 16.3K to 4.2K. Overall, the full configuration achieves the best balance, covering all 68 true positives with only 6 false positives, 4.2K tokens, and 31.2s per project.

\begin{table}[htbp]
\centering
\caption{Results of the ablation study.}
\label{tab:rq2-ablation}
\setlength{\tabcolsep}{3pt}
\resizebox{\linewidth}{!}{
\begin{threeparttable}
\begin{tabular}{lrrrrrr}
\hline
\textbf{Variant} & \textbf{\#Static Cand.} & \textbf{\#Alerts} & \textbf{\#TP} & \textbf{\#FP} & \textbf{Avg. Tok.} & \textbf{Avg. T} \\
\hline
Full \toolname{} & 340 & 74 & \cellcolor{best}\textbf{68} & \cellcolor{best}\textbf{6} & \cellcolor{best}\textbf{4.2K} & 31.2s \\
\emph{w/o Framework Modeling} & 910 & 276 & 61 & 215 & 22.3K & 92.3s \\
\emph{w/o Agentic Gate} & 1,453 & 87 & 62 & 25 & 40.4K & 110.3s \\
\emph{w/o Bound Coverage} & 365 & 70 & 60 & 10 & 4.3K & 38.1s \\
\emph{w/o Benign Loop Filtering} & 696 & 103 & 62 & 41 & 16.3K & 66.3s \\
\emph{w/o LLM Pruning} & 340 & 340 & \cellcolor{best}\textbf{68} & 272 & 0 & \cellcolor{best}\textbf{3.8s} \\
\hline
\end{tabular}
\begin{tablenotes}[flushleft]
\footnotesize
\item \textit{Note.}  The evaluation subset contains 264 agent projects. \textbf{Avg. Tok.} and \textbf{Avg. T} denote the average token usage and analysis time per project.
\end{tablenotes}
\end{threeparttable}
}
\end{table}

\subsubsection{Baseline Comparison}

There is no existing tool dedicated to detecting IAL failures. We therefore compare \toolname{} with two LLM-based baselines: a general coding agent and a pure LLM API analysis. This comparison examines whether IAL detection can be replaced by a coding agent or direct LLM prompting, given the broad use of LLMs in vulnerability detection and risk analysis.
All three methods use \texttt{gpt-5.5} as the base model and a 180s timeout per project. The coding agent baseline uses Codex with no internet access and a prompt that asks it to analyze the entire repository. The pure LLM API baseline uses the same core detection prompt but analyzes Python files one by one. We include the complete prompts in our artifacts. To control cost and time, we limit each project to at most 80 Python files and each file to at most 3,000 characters. In the 264 project evaluation subset, this baseline reads 12,958 of 34,634 Python files, including 9,233 files read in full.
\autoref{tab:rq2-baselines} shows that general LLM baselines are not a substitute for \toolname{}. The pure LLM API baseline covers only 23 of the 68 confirmed failures and produces 183 alerts. Although it fully reads only about one quarter of all Python files, its token usage is already more than four times that of \toolname{}. The coding agent baseline has higher recall than the pure LLM API baseline, covering 50 confirmed failures, but it produces many extra alerts, and reaches the timeout or error limit in 75 projects. It also incurs much higher cost, with 141.86K tokens and 116.0s per project on average. Due to cost constraints, we bound the analysis time and input size for the LLM baselines. Even so, the results show that \toolname{} achieves higher coverage and precision, lower token usage, and shorter analysis time.

\begin{table}[htbp]
\centering
\caption{Comparison with LLM-based baselines.}
\label{tab:rq2-baselines}
\setlength{\tabcolsep}{3.5pt}
\resizebox{\linewidth}{!}{
\begin{threeparttable}
\begin{tabular}{lrrrrrr}
\hline
\textbf{Method} & \textbf{\#Alerts} & \textbf{\#TP Cov.}  & \textbf{\#Missed} & \textbf{\#Timeout} & \textbf{Avg. Tok.} & \textbf{Avg. T} \\
\hline
\toolname{} & \cellcolor{best}\textbf{74} & \cellcolor{best}\textbf{68} & \cellcolor{best}\textbf{0} & \cellcolor{best}\textbf{0} & \cellcolor{best}\textbf{4.2K} & \cellcolor{best}\textbf{31.2s} \\
Coding assistant & 140 & 50 & 18 & 75 & 141.9K & 116.0s \\
Pure LLM API & 183 & 23 & 45 & 1 & 18.1K & 34.4s \\
\hline
\end{tabular}
\begin{tablenotes}[flushleft]
\footnotesize
\item \textit{Note.}  Results are measured on the evaluation subset of 264 agent projects. The coding assistant baseline uses Codex. All methods use \texttt{gpt-5.5} as the base model and a 180s timeout per project. \#TP Cov. denotes the number of confirmed IAL failures covered by each method.
\end{tablenotes}
\end{threeparttable}
}
\end{table}

\subsubsection{FP \& FN}
For the 6 false positives from the real-world detection in RQ1, we inspected each case in detail. Each case contains a real agentic feedback path, but manual review found an effective bound covering the path. These bounds are difficult to resolve statically because they are indirect or framework dependent: two workflows use \texttt{max\_steps} or fallback counters outside the loop body, one OpenAI Agents supervisor relies on the default \texttt{max\_iterations=3}, one LangGraph tool loop is capped by \texttt{max\_tool\_calls=2}, one QA workflow combines a retry budget with asyncio timeouts, and one AutoGen classification flow is a bounded two turn exchange. These cases show that most false positives come from subtle bound configurations rather than spurious loop matches. \toolname{} conservatively retains them to avoid missing IAL failures with similar feedback structures.

To estimate false negatives, we reviewed baseline alerts not reported by \toolname{}. The two baselines produced 250 raw extra alerts. After removing two alerts already matched to \toolname{} findings and deduplicating overlapping alerts at the same source location, 239 unique baseline-only locations remained. Two authors independently labeled them with 92.1\% agreement, and all disagreements were resolved with a third author. This process confirmed 7 false negatives of \toolname{}.
These false negatives mainly involve LangGraph workflow or tool cycles with unrecognized bounds, handwritten tool call loops, framework-adjacent orchestration loops, and multi-agent conversations missing effective turn or termination bounds. Most other baseline-only alerts are not IAL failures, such as tutorials, examples, human-driven REPLs, UI or service lifecycle loops, bounded workflows, or reports without enough evidence of an autonomous agent feedback path.

\subsection{RQ3: Stability}
\label{sec:rq3-stability}

\subsubsection{End-to-end Repeatability}
We repeat the end-to-end analysis three times on the evaluation subset using the same static configuration and default LLM-assisted pruning. As shown in \autoref{tab:rq3-static-repeat}, the static stage is fully repeatable: all runs produce the same 340 static candidates. The variation comes from LLM pruning, which reports 74, 73, and 70 alerts and covers 68, 64, and 60 true positives, respectively. Token cost remains stable at 4.2K tokens per project, while average time ranges from 31.2s to 39.3s.
These results show that static candidate discovery is deterministic, whereas LLM pruning introduces limited but visible variability. We therefore use the LLM only as a pruning aid, and base the RQ1 result on manual review of the 74 potential findings from the default run.

\begin{table}[htbp]
\centering
\caption{Results of end-to-end repeatability experiments.}
\label{tab:rq3-static-repeat}
\setlength{\tabcolsep}{5pt}
\resizebox{\linewidth}{!}{
\begin{tabular}{lrrrrr}
\hline
\textbf{Run} & \textbf{\#Static Cand.} & \textbf{\#Alerts} & \textbf{\#TP Cov.} & \textbf{Avg. Tok.} & \textbf{Avg. T} \\
\hline
Run 1 & 340 & 74 & 68 & 4.2K & 31.2s \\
Run 2 & 340 & 73 & 64 & 4.2K & 39.3s \\
Run 3 & 340 & 70 & 60 & 4.2K & 38.6s \\
\hline
\end{tabular}}
\end{table}

\subsubsection{LLM Sensitivity}

We evaluate LLM sensitivity to clarify the role of the LLM-assisted pruning stage. The input is fixed to the same 340 static candidates, so this experiment only measures how different models filter the same static evidence, rather than how candidates are discovered.
As shown in \autoref{tab:rq3-llm-stability}, pruning behavior varies noticeably across models. The default setting, \texttt{gpt-5.5}, reduces 340 candidates to 74 alerts while covering all 68 confirmed true positives from RQ1. In contrast, \texttt{gpt-5.4-mini} and \texttt{gemini-2.5-flash} keep more alerts, 136 and 131 respectively, but cover fewer true positives. This shows that retaining more candidates does not necessarily improve coverage; the model must correctly interpret whether the feedback path is feasible and effectively bounded. \texttt{deepseek-v4-pro} reports a similar number of alerts to the default model, but covers fewer true positives with higher token and time cost.

\begin{table}[htbp]
\centering
\caption{LLM-assisted pruning sensitivity.}
\label{tab:rq3-llm-stability}

\resizebox{\linewidth}{!}{
\begin{tabular}{lrrrr}
\hline
\textbf{Run} & \textbf{\#Alerts} & \textbf{\#TP Cov.} & \textbf{Avg. Tok.} & \textbf{Avg. T} \\
\hline
\rowcolor{gray!8}
\multicolumn{5}{l}{\textbf{Default Model}} \\ \hline
\texttt{gpt-5.5}  & 74 & 68 & 4.2K & 31.2s \\ \hline
\rowcolor{gray!8}
\multicolumn{5}{l}{\textbf{Model Variants}} \\ \hline
\texttt{gpt-5.4-mini}  & 136 & 41 & 3.8K & 7.4s \\
\texttt{deepseek-v4-pro}  & 78 & 54  & 6.2K & 48.2s \\
\texttt{gemini-2.5-flash}  & 131 & 47 & 7.0K & 17.8s \\
\hline
\end{tabular}}
\end{table}

\section{Discussion}
\label{sec:discussion}

\subsection{Implications}
Our findings suggest that IAL prevention should be considered in both framework design and application practice. For framework developers, bounds should be enforced at the runtime scope where feedback is created, rather than exposed only as optional local parameters. When graph transitions, tool dispatch, conversation management, or handoffs can reenter model or agent execution, the framework should provide default budgets, propagate them across the feedback path, and report uncovered cycles during compilation or runtime. Frameworks should also expose guards for state growth, since repeated message or workflow updates can enlarge the context even before execution limits are reached.
For framework users, normal agent iteration should be deployed with explicit stopping rules. Developers should not rely on the model to eventually stop producing tool calls, valid plans, or termination messages. Each agent run should set turn or step limits, retry and repair paths should have caps and timeouts, and message history or workflow state should have size limits. 

\subsection{Limitations}
\toolname{} has several limitations. First, as a static analyzer, it over-approximates possible agent-specific dependencies rather than predicting a single execution trace and may report false positives. Second, the current implementation of \toolname{} focuses on Python applications built with eight agent frameworks, so unsupported languages, frameworks, or incomplete framework models may lead to false negatives. Third, \toolname{} has limited support for highly customized user-defined semantics, such as project-specific schedulers, external-state-based stopping logic, or semantic checks over natural-language outputs, which may cause imprecision in bound reasoning. Fourth, \toolname{} uses LLM-assisted pruning only as an optional negative filter, but LLM judgments may still be incomplete or unstable for ambiguous cases.


\section{Conclusion}
\label{sec:conclusion}

In this paper, we studied IALs, a structural execution failure in which agentic feedback paths repeatedly trigger model, tool, agent, or workflow execution without an effective termination condition. To detect such failures before deployment, we presented \toolname{}, a static analyzer that normalizes source code and framework behavior into Agent IR, constructs an ALDG, and identifies feedback paths that can repeatedly reach costly or state-growing actions without effective bound coverage. Our evaluation on 6,549 real-world repositories demonstrates that \toolname{} can uncover practical IAL failures across diverse agent projects with high precision. These results highlight the need for agent-aware static analysis to make iterative agent execution safer and more controllable.

\section*{Acknowledgements}

We used ChatGPT and Codex to assist with language polishing and code refinement. All ideas, approaches, experimental designs, and results are the authors' own work.

\newpage
\balance
\bibliographystyle{IEEEtranS}
\bibliography{main}

\end{document}